\documentclass{PoS}
\pdfoutput=1
\usepackage{amsmath}
\usepackage{mathtools}
\usepackage{graphicx}

\usepackage[utf8]{inputenc}

\title{Chiral phase transition of (2 + 1)-flavor QCD on $N_{\tau} = 6$ lattices}

\ShortTitle{Chiral phase transition of (2 + 1)-flavor QCD}

\author{\speaker{Sheng-Tai Li}\thanks{The numerical simulations were carried out on clusters of
		the USQCD Collaboration in Jefferson Lab and Fermilab, TianHe I \& II supercomputer centers in China and on the local computing cluster at Central China Normal University. The anthors are partly supported 
		by the National Natural Science Foundation of China under grant numbers 11535012 and 11521064.} \ and Heng-Tong Ding (for the Bielefeld-BNL-CCNU collaboration)\\
        Key Laboratory of Quark \& Lepton Physics (MOE) and Institute of Particle Physics,  \\
        Central China Normal University, Wuhan 430079, China.\\
        E-mail: \email{lishengtai@mails.ccnu.edu.cn \&
        	hengtong.ding@mail.ccnu.edu.cn}}%

\abstract{We present updated studies on the chiral phase transition in $N_{f}=2+1$ QCD. Simulations have been carried out using Highly Improved Staggered Quarks (HISQ) on lattices with temporal extent $N_{\tau} = 6$ at vanishing baryon chemical potential. We updated our previous study \cite{Ding:2015pmg} by extending the temperature window from (140 MeV, 150 MeV) to (140 MeV, 170 MeV). The strange quark mass was chosen to its physical value $m_{s}^{\mathrm{phy}}$, and five values of two degenerate light quark masses are varied from $m_{s}^{\mathrm{phy}}/80$ to $m_{s}^{\mathrm{phy}}/20$ which correspond to a Goldstone pion mass ranging from 80 MeV to 160 MeV in the continuum limit. The universal scaling behaviour of the QCD chiral phase transition is investigated by studying the temperature and quark mass dependences of chiral condensates and chiral susceptibilities. The window of criticality compared to previous studies is also discussed.	}

\FullConference{34th annual International Symposium on Lattice Field Theory\\
	24-30 July 2016\\
	University of Southampton, UK}

\begin{document}

\section{Introduction}
\begin{figure}[htp]
	\begin{center}
	\includegraphics[width=0.35\textwidth]{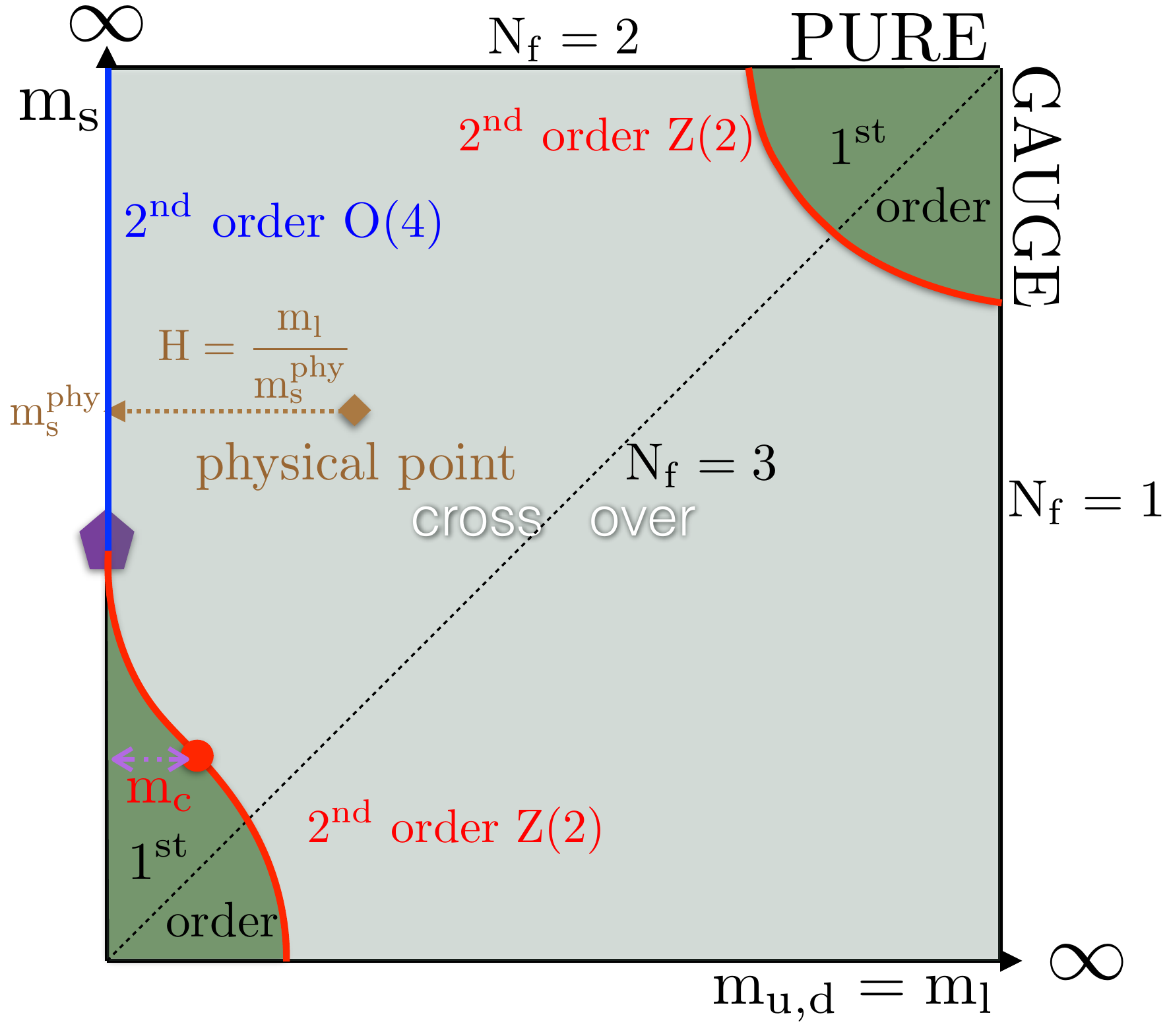}
	\includegraphics[width=0.35\textwidth]{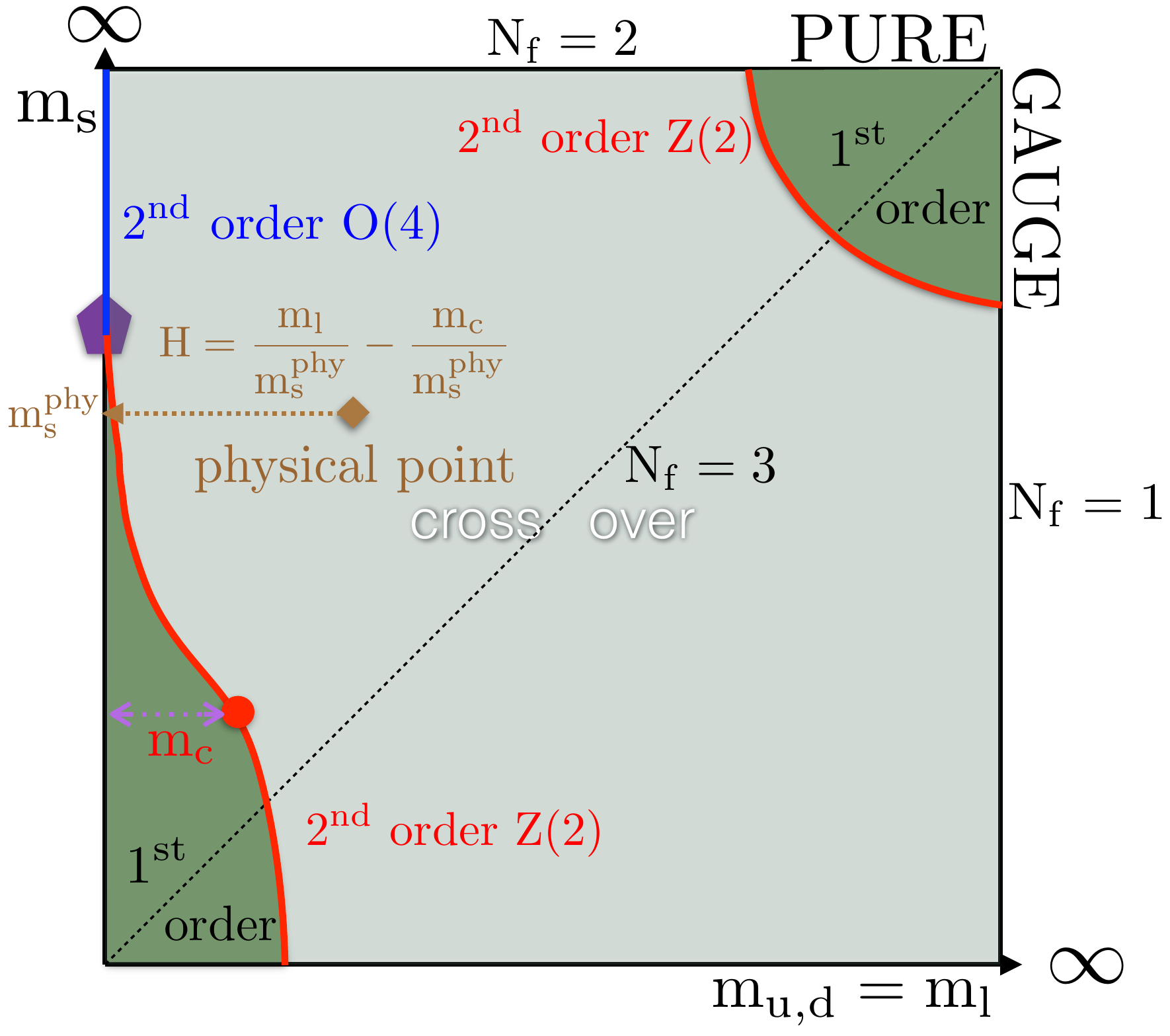}~
	\end{center}
	\caption{Schematic QCD phase structure with different values of quark masses ($m_{u,d},m_{s}$) at zero baryon number density for $m_{s}^{\mathrm{tri}}<m_{s}^{\mathrm{phy}}$ (left) and $m_{s}^{\mathrm{tri}}>m_{s}^{\mathrm{phy}}$ (right).}
\label{Fig:latticedata}	
\end{figure}

Understanding the QCD phase diagram is one of the basic goals of lattice QCD calculations at non-zero temperature.
The QCD phase structure may depend on the number of light quark flavors \cite{PhysRevD.29.338}.
The order of the phase transitions as a function of quark mass at zero baryon chemical potential is summarized in Fig.~\ref{Fig:latticedata}.
The physical point ($m_{u}^{\mathrm{phy}},m_{s}^{\mathrm{phy}}$) is confirmed to be in the crossover region~\cite{Bazavov:2011nk,Aoki:2006we}.
For three degenerate flavors of quarks, $N_{f}=3$, first order phase transitions are observed in the limits of both infinitely large and small masses~\cite{Ding:2015ona,Schmidt:2017bjt}.
This corresponds to deconfinement and chiral phase transitions as shown in the upper right and bottom left corners of Fig.~\ref{Fig:latticedata}, respectively. 
The first order phase transition regions and the crossover region are separated by second order phase transition lines which belong to the Z(2) universality class. 
In the chiral limit of $N_{f}=2$ theory, if $U_{A}(1)$ symmetry is broken, the chiral phase transition is a second order phase transition belonging to an O(4) universality class~\cite{PhysRevD.29.338}.
Thus the chiral first order region for the $N_{f}=3$ case, the second order O(4) line for $N_{f}=2$ case and the second order Z(2) line are supposed to meet in a tri-critical point $m_{s}^{\mathrm{tri}}$. 
The location of the tri-critical point is uncertain, even it is possible that the tri-critical point shifts to infinite strange quark mass~\cite{Philipsen:2016hkv}.

It is not yet clear whether the order of phase transition is first or second for 2-flavor QCD in the chiral limit.
If $m_{s}^{\mathrm{tri}}<m_{s}^{\mathrm{phy}}$, it is expected that in the chiral limit there will be a second order phase transition which belongs to the O(4) universality class as seen from the Fig.~\ref{Fig:latticedata} (left).
If $m_{s}^{\mathrm{tri}}>m_{s}^{\mathrm{phy}}$, towards the chiral limit the system passes through the Z(2) critical line to a first order phase transition region as shown in Fig.~\ref{Fig:latticedata} (right).
Since it is impossible to perform a lattice simulation in a true chiral limit, it is thus not so surprising that there is no conclusion yet~\cite{Ding:2015ona,Schmidt:2017bjt}.

The Highly Improved Staggered Quark (HISQ) action is expected to reduce the lattice artifacts caused by the taste-symmetry breaking by more than an order of magnitude compared to the p4fat3~\cite{Bazavov:2011nk}.
In this proceedings, we report our current state of the investigation on the universal behaviour of chiral phase transition on $N_{\tau}=6$ lattices with HISQ action in a larger temperature window 140 MeV$<T<$170 MeV compared to our previous study~\cite{Ding:2015pmg}.
\section{Universal scaling behaviour of chiral phase transition}
The universal behavior of the order parameter $M$ and its susceptibility $\chi_{M}$ can be described by the so-called Magnetic Equation of State (MEOS) as follows
\begin{small}
\begin{equation}
M(t,h)=h^{1/\delta}f_{G}(z)\quad \mathrm{and}\quad
\chi_M(t,h) = \frac{\partial M}{\partial H} =  h_{0}^{-1}h^{1/\delta-1}f_{\chi}(z).
\label{eq.M}
\end{equation}
\end{small}
Here $z=th^{-1/\beta\delta}$ is the scaling variable, $t=\frac{1}{t_{0}}\frac{T-T_{c}}{T_{c}}$ is a reduced temperature and $h=H/h_{0}=\frac{m_{l}}{m_{s}}/h_{0}$ is the symmetry breaking field. The $\beta$, $\delta$ are universal critical exponents which only rely on the symmetry of the system, 
and the three non-universal parameters $h_{0}$, $t_{0}$, $T_{c}$ are unique for a particular system, e.g. $T_{c}$ is the critical temperature.
These non-universal parameters can be determined by studying the scaling behaviours of the chiral order parameter $M$ and its susceptibility $\chi_{M}$ as expressed in Eq.~(\ref{eq.M}). The scaling function $f_{\chi}$ is $\frac{1}{\delta}(f_{G}(z)-f_{G}^{'}(z)z/\beta$),
and the peak location of $f_{\chi}$ is characterized by $z_{p}$. 
It is related to the pseudo-critical temperature $T_{pc}$ as follows
\begin{small}
\begin{equation}
T_{pc}(H)=T_{c}\frac{z_{p}}{z_{0}}H^{1/\beta\delta}+T_{c},
\label{eq.M1}
\end{equation}
\end{small}
where the non-universal parameter $z_{0}=h_{0}^{1/\beta\delta}/t_{0}$ remains unchanged with the rescaling of the order parameter $M$~\cite{Ejiri:2009ac}. 
With the parameter $z_{0}$, the scaling variable z thus can be written as,
\begin{small}
\begin{equation}
z=th^{-1/\beta\delta}=z_{0}\frac{T-T_{c}}{T_{c}}H^{-1/\beta\delta}.
\label{eq.M2}
\end{equation}
\end{small}
This indicates that a change of the scaling variable $\Delta z$ is proportional to a change of temperature $\Delta T$ for a fixed quark mass.

In the chiral limit of two quark flavors the $N_{f}=2$ QCD Lagrangian has a $SU_{L}(2)\times SU_{R}(2)$ chiral symmetry which is isomorphic to O(4).
If the physical point is above the tri-critical point, $m_{s}^{\mathrm{phy}} > m_{s}^{\mathrm{tri}}$, for the case of $N_{f}=2+1$ QCD in the chiral limit of the two light quarks with its strange quark mass fixed at the physical point,
it is expected that the chiral phase transition belongs to the same universality class as 3-dimensional O(4) symmetric spin models.
The remnant chiral symmetry of the HISQ action on the lattice is isomorphic to O(2) at nonzero lattice spacing.
This symmetry gives rise to one rather than three massless Goldstone modes in the chiral limit.
Since the critical exponents and scaling functions belonging to the O(2) and O(4) universality classes are similar, we hope to observe generic evidence for the O(N) scaling from our lattice calculation using the HISQ action.

It may happen that the physical point is below the tri-critical point.
We do not know the boundary of the chiral first order region for (2 + 1)-flavor QCD.
The chiral first order phase transition is supposed to end at the second order phase transition line i.e. at the critical quark mass $m_{c}$.
Here $m_{c}$ is the critical quark mass (not to be confused with the charm quark mass).
The breaking field $H=\frac{m_{l}}{m_{s}}$ is then replaced by $\frac{m_{l}}{m_{s}}-r_{c}$, where $r_{c}=\frac{m_{c}}{m_{s}}$.
To determine the value of $r_{c}$
we will need to perform a Z(2) MEOS fit to the chiral order parameter $M$. 
A vanishing value of $r_{c}$ within error thus suggests that $m_{s}^{\mathrm{phy}} > m_{s}^{\mathrm{tri}}$.
     
\section{Lattice setup}
We perform lattice simulations using Highly Improved Staggered Quarks and tree-level improved gauge action (HISQ/tree) on $N_{\tau}=$ 6 lattices.
The strange quark mass is chosen to its physical quark mass value $m_{s}^{\mathrm{phy}}$, and five values of light quark masses are taken as $m_{s}^{\mathrm{phy}}/20$, $m_{s}^{\mathrm{phy}}/27$, $m_{s}^{\mathrm{phy}}/40$, $m_{s}^{\mathrm{phy}}/60$, $m_{s}^{\mathrm{phy}}/80$,
whose corresponding lattice size, $N_{\sigma}$, are $24$, $24$, $32$, $40$, $32$, respectively to ensure $m_{\pi}L \gtrsim 4$. 
For our lightest light quark $m_{l}=m_{s}^{\mathrm{phy}}/80$, the simulations are also performed on $N_{\sigma}=$48. 

For each data set, we have generated about 1500 gauge field configurations separated by 10 trajectories.
We used 1024 noise vectors on each gauge field configuration and constructed unbiased estimators for the various traces to compute the chiral condensate and its susceptibility.
\section{Results}
The main contribution of the divergent term is linear in the quark mass.
To suppress the UV divergent, we study the subtracted chiral order parameter $M$ and its susceptibility $\chi_{M}$, they are expressed in units of $T^{4}$ to be dimensionless as follows
\begin{small}
\begin{equation}
		M={m_{s}}\left(\left\langle\bar{\psi}\psi\right\rangle_{l} -\frac{2m_{l}}{m_{s}}\left\langle\bar{\psi}\psi\right\rangle_{s}\right)/T^{4},
\label{eq.sub_pbp}
\end{equation}
\end{small}
\begin{small}
\begin{equation}
\chi_{M}\equiv\frac{\partial M}{\partial H}  \equiv{m_{s}}^2\chi_{l,\mathrm{subtot}}/T^{4}, \quad  \chi_{l,\mathrm{subtot}}=\frac{\partial }{\partial m_{l}}\left(\left\langle\bar{\psi}\psi\right\rangle_{l} -\frac{2m_{l}}{m_{s}}\left\langle\bar{\psi}\psi\right\rangle_{s}\right).
\label{eq.sub_pbp2}
\end{equation}
\end{small}
\begin{figure}[ht]
	\begin{center}
		\includegraphics[width=0.41\linewidth]{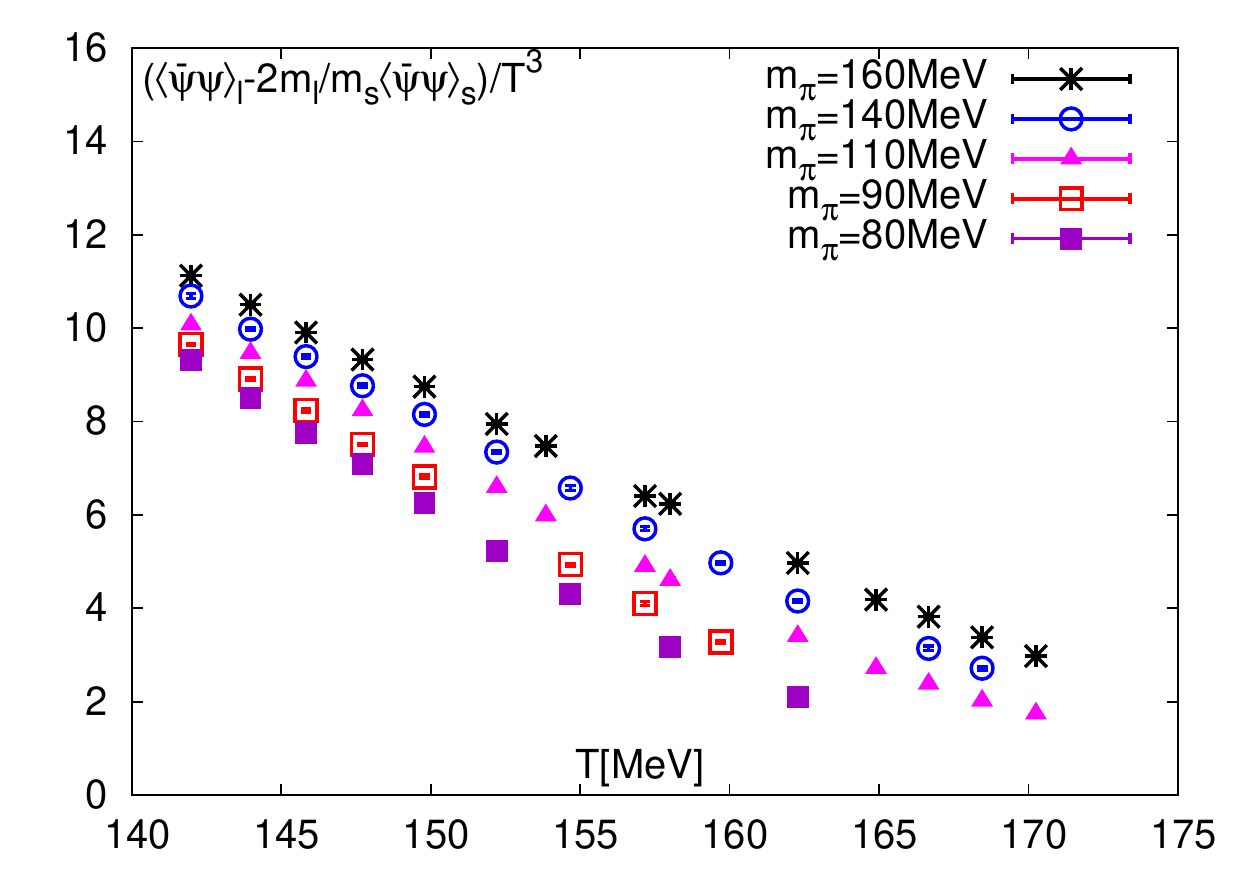}
		\includegraphics[width=0.41\linewidth]{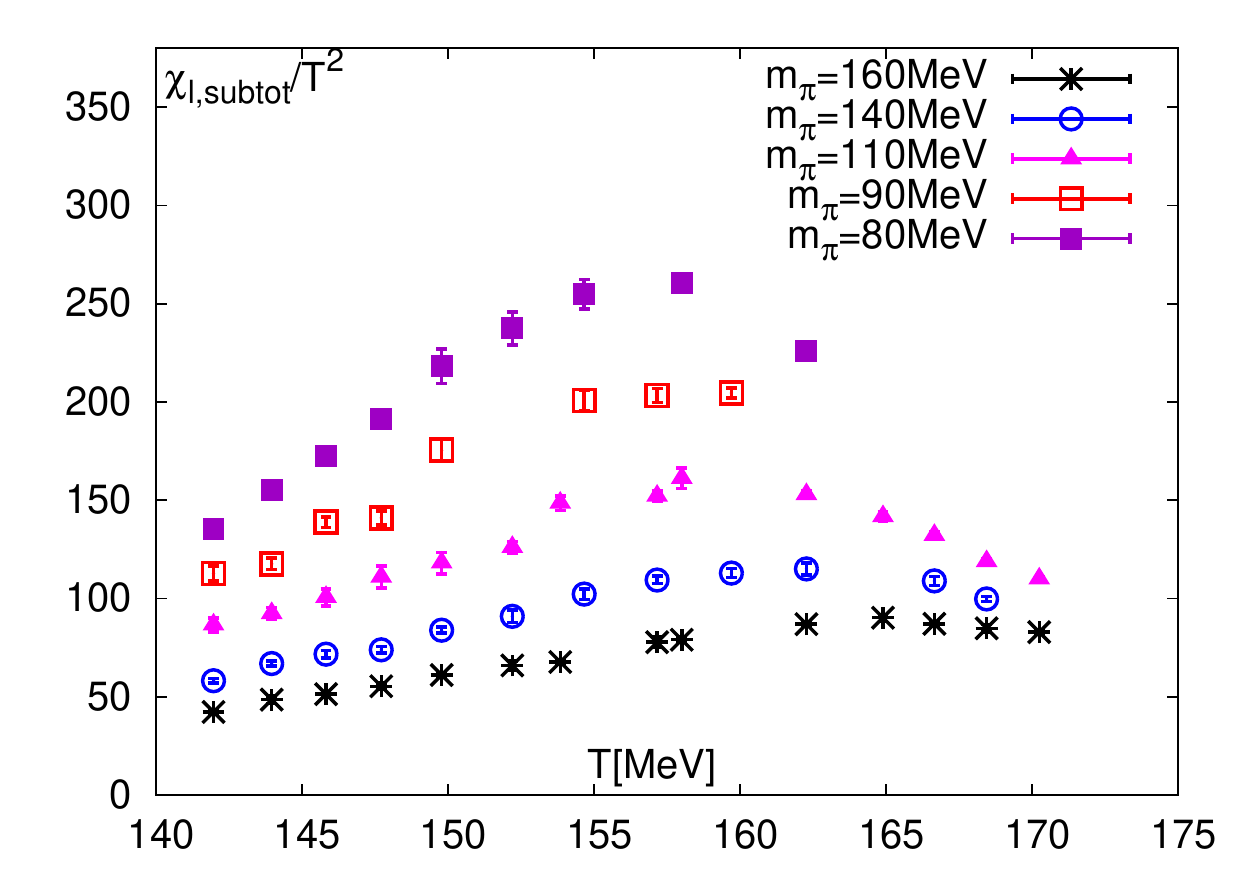}	
	\end{center}
		\caption{
Quark mass and temperature dependences of the subtracted chiral condensates (left) and their susceptibilities (right). 
		}
\label{Fig:data_sets}
\end{figure}

As shown in Fig.~\ref{Fig:data_sets}, the subtracted chiral condensates decrease with decreasing $m_{\pi}$ at a fixed value of the temperature, and also decrease with increasing temperature at a fixed pion mass $m_{\pi}$. 
For their susceptibilities $\chi_{l,\mathrm{subtot}}$, the positions of the peaks move to lower temperatures  with decreasing light quark mass, which is consistent with Eq.~(\ref{eq.M1}).
The height of the peak increases at smaller light quarks which suggests that the system approaches to a phase transition in the chiral limit.
\begin{figure}[ht]
	\begin{center}
		\includegraphics[width=0.329\linewidth]{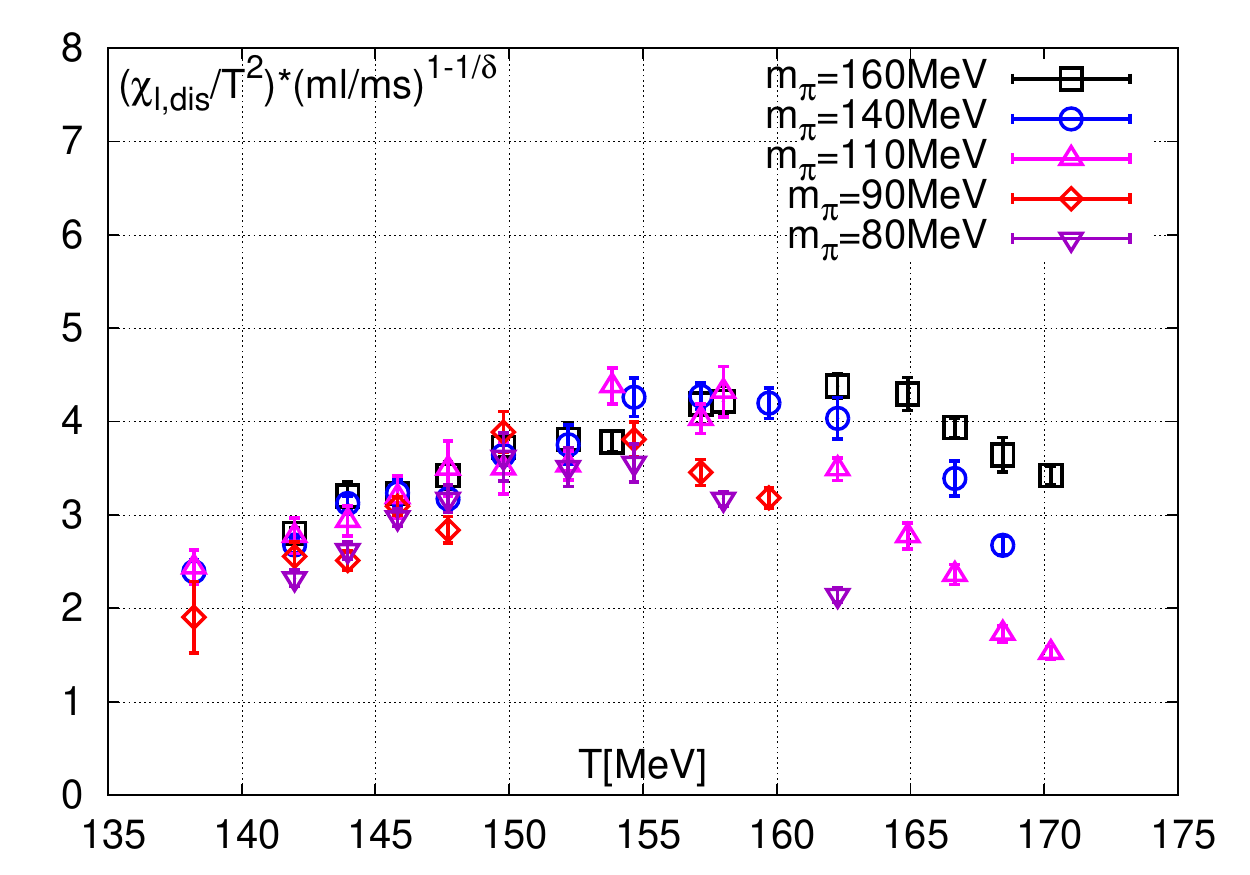}	
		\includegraphics[width=0.329\linewidth]{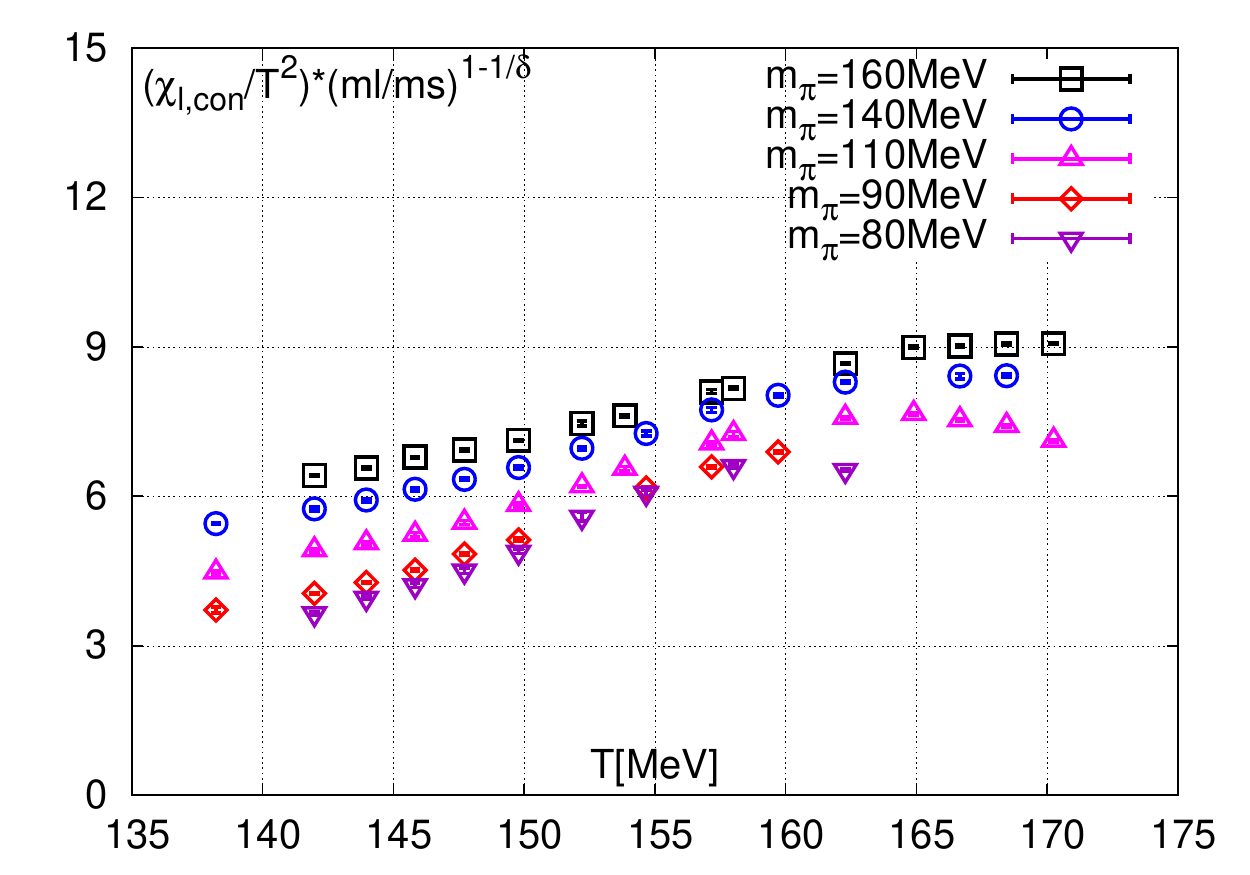}
		\includegraphics[width=0.329\linewidth]{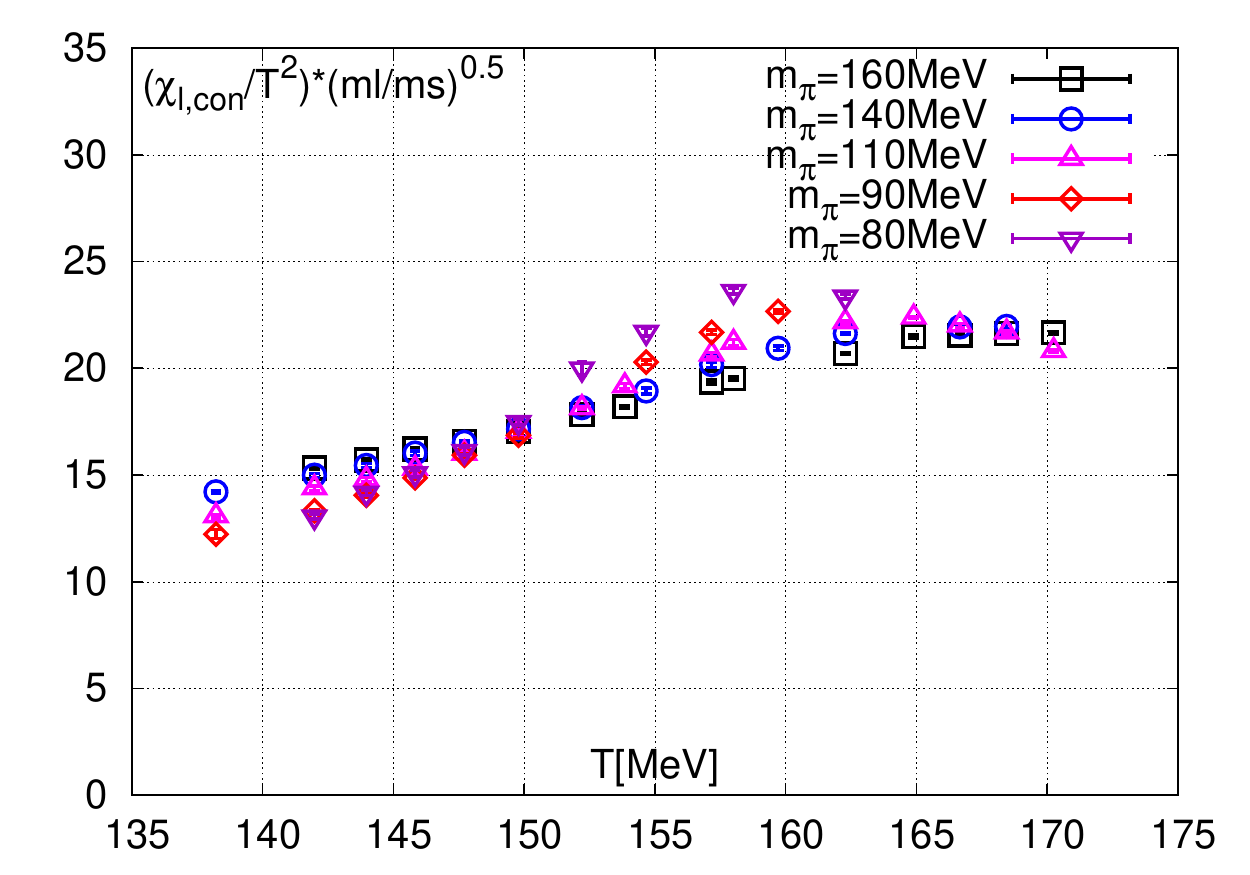}
	\end{center}
	\caption{The disconnected part of the chiral susceptibility multiplied by $(m_{l}/m_{s})^{1-1/\delta}$ (left). The connected part of chiral susceptibility multiplied by $(m_{l}/m_{s})^{1-1/\delta}$ (middle) and $(m_{l}/m_{s})^{0.5}$ (right). Here $\delta = 4.780$ is a critical exponent belonging to the O(2) universality class.
	}
	\label{Fig:conn}
\end{figure}

We also study the temperature and quark mass dependences of the connected and disconnected parts of the chiral susceptibility,
\begin{small}
\begin{equation}
\chi_{l,dis}=\frac{1}{4N_{\sigma}^{3}N_{\tau}}\left[\left\langle(\mathrm{Tr}M_{l}^{-1})^{2}\right\rangle-\left\langle \mathrm{Tr}M_{l}^{-1}\right\rangle^{2}\right],\quad \chi_{l,con}=-\frac{1}{2}\mathrm{Tr}\sum_{x}\left\langle M_{l}^{-1}(x,0)M_{l}^{-1}(0,x)\right\rangle.
\label{eq.sus}
\end{equation}
\end{small}
As seen from the left plot of Fig.~\ref{Fig:conn}, the disconnected susceptibility is roughly proportional to $m_{l}^{1/\delta -1}$ in the temperature range 140 MeV$<T<$150 MeV.
While the connected susceptibility is approximately proportional to $m_{l}^{-0.5}$ in the whole temperature range. 
The disconnected susceptibility is the fluctuation of the chiral condensate and corresponds to fluctuations of the order parameter in the O(N) spin models. Thus it is expected to have a large contribution to the critical behaviour of the QCD system.
However, this is not the case for the connected susceptibility.
The connected susceptibility with large light quark mass is responsible for additional scaling violations other than the scaling function $f_{\chi}$.
This may explain the scaling violation which appeared in the $f_{\chi}$ in our previous study~\cite{Ding:2015pmg}.
Note that the connected susceptibility scaled by the dominant singularity $(m_l/m_s)^{1-1/\delta}$ significantly decreases with quark mass in the whole temperature window as seen in the middle plot of Fig.~\ref{Fig:conn}.
Since $1-1/\delta$ is larger than 0.5, $\chi_{l,con}/H^{1/\delta-1}$ has a positive quark mass power, it vanishes in the chiral limit and then has no contribution to the singular behaviour of the chiral phase transition. 

\begin{figure}[ht]
	\begin{center}
		\includegraphics[width=0.329\linewidth]{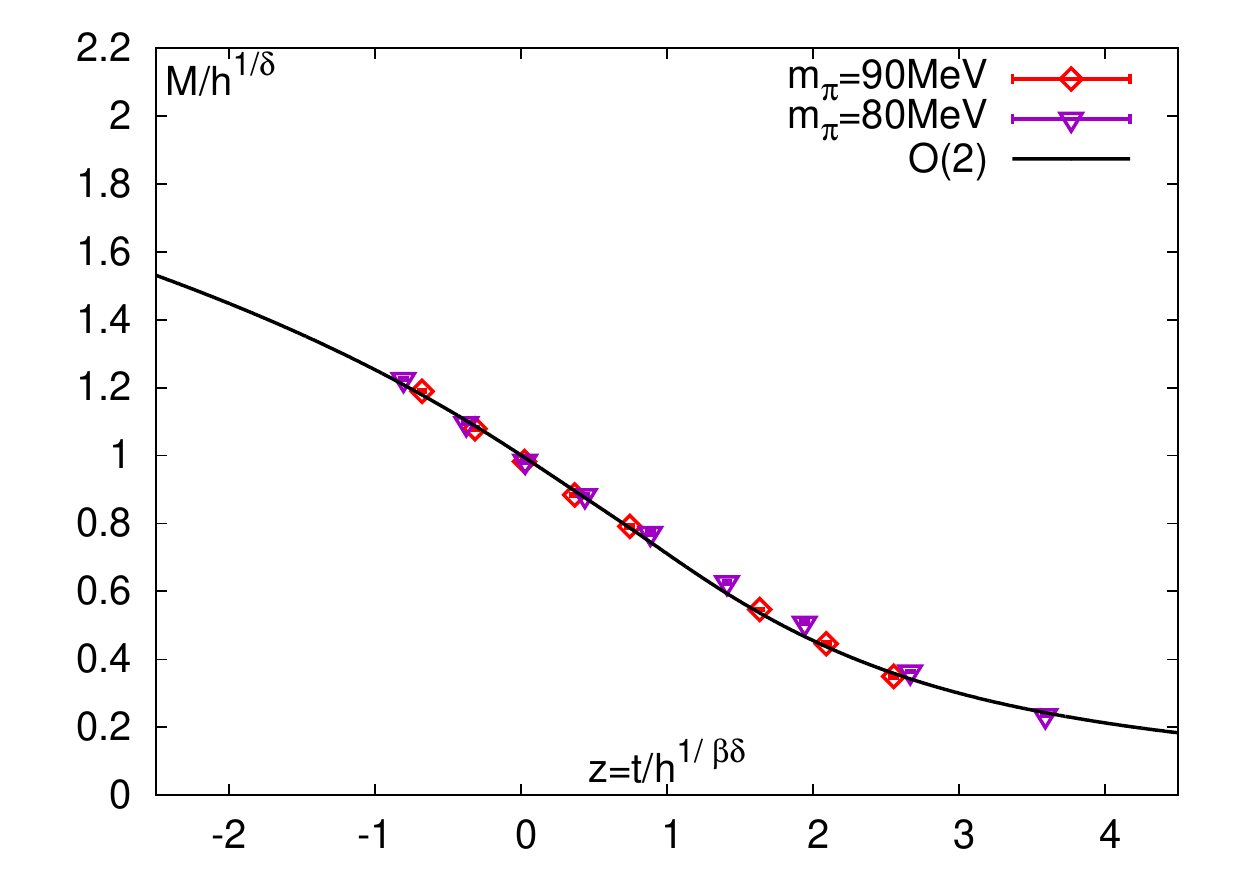}
		\includegraphics[width=0.329\linewidth]{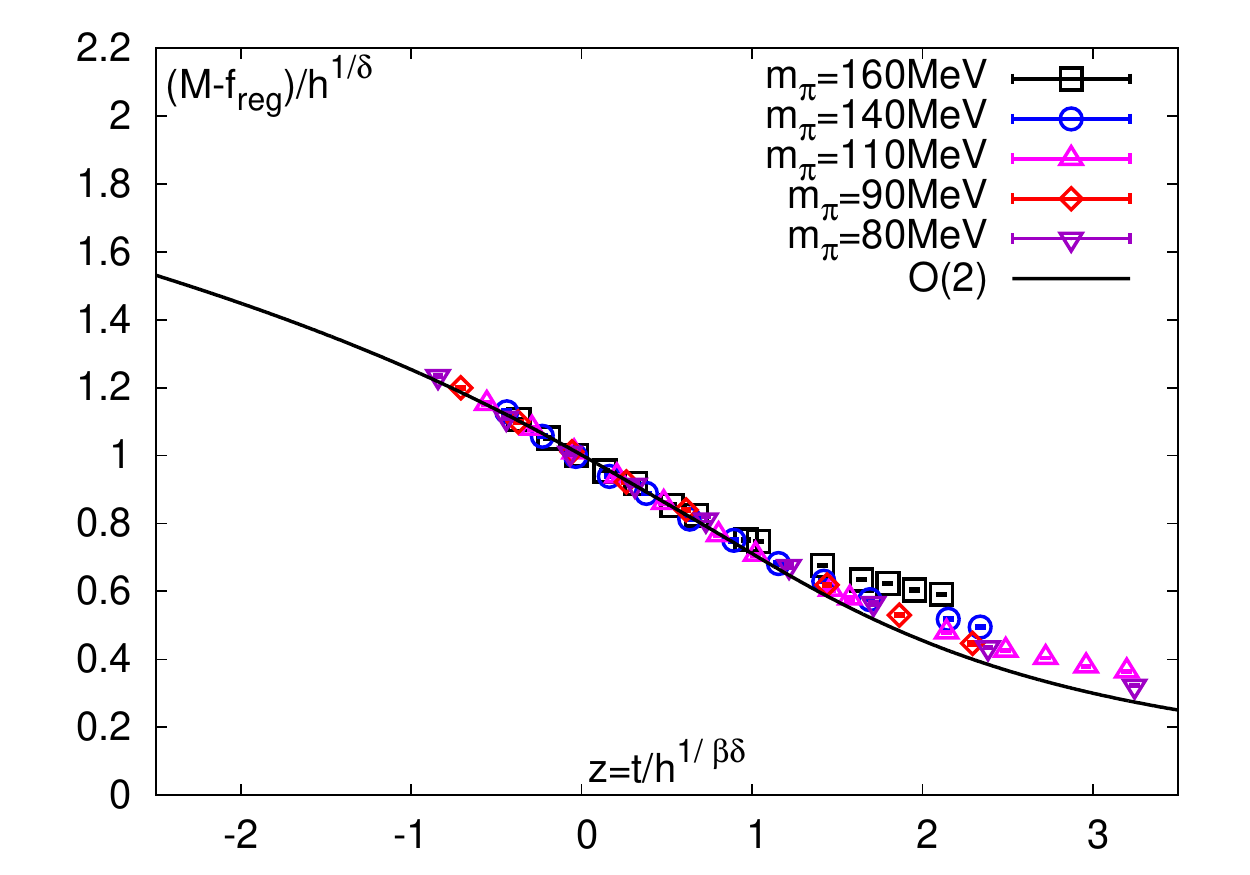}
		\includegraphics[width=0.329\linewidth]{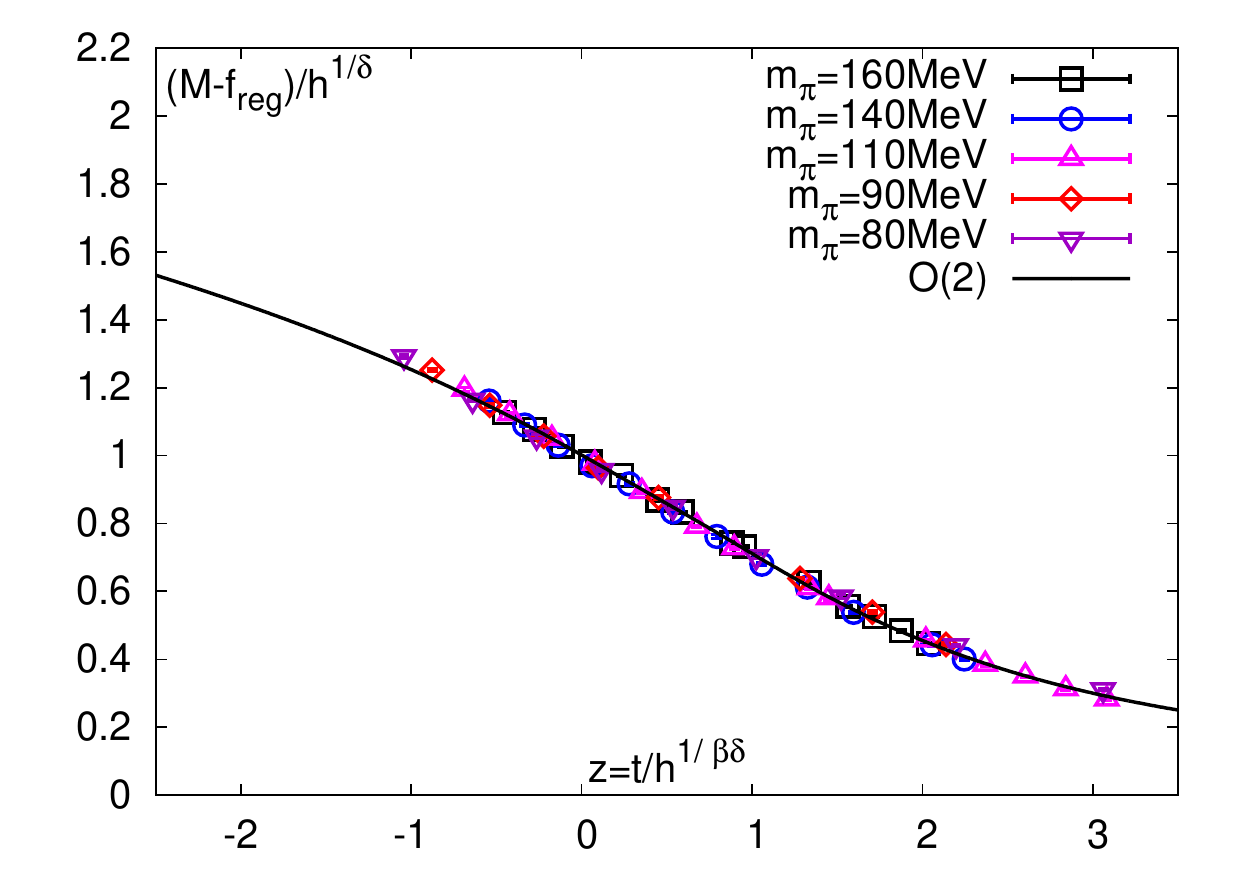}
	\end{center}
		\caption{Left: We perform a MEOS fit to $M$ using $m_{l}=m_{s}/80$
        and $m_{l}=m_{s}/60$ data sets without a regular part in a
        temperature range 140 MeV$<T<$150 MeV, and then we use the
        obtained parameters $t_{0}$, $h_{0}$, $T_{c}$ to replot the data
        sets in the $\textit{whole}$ temperature range 140 MeV$<T<$162
        MeV. Middle: We perform a MEOS fit to $M$ using $\textit{all}$ the
        data sets in a temperature range 140 MeV$<T<$150 MeV with a
        regular term
        $f_{\mathrm{reg}}=(a_{0}+a_{1}\frac{T-T_{c}}{T_{c}})\frac{m_{l}}{m_{s}}$,
        then we use the obtained parameters to replot $\textit{all}$ the
        data sets in the $\textit{whole}$ temperature range 140
        MeV$<T<$170 MeV. Right: We perform a MEOS fit to $M$ using
        $\textit{all}$ the data sets. The used regular term includes a
        higher order term in $T$ as expressed in Eq.~\protect\eqref{eq.fit}.
    }
\label{Fig:con}			 
\end{figure}

In the following we will focus on the scaling analysis of the chiral order parameter $M$.
We present a MEOS fit according to Eq.~(\ref{eq.M}) to the subtracted chiral condensate 
in a temperature range 140 MeV$<T<$150 MeV and with the $m_{\pi}$ = 80, 90 MeV data sets only.
To check the scaling window in the $T$ direction,
 we then use the obtained fit parameters ($t_{0},\ h_{0},\ T_{c}$) to replot the data sets in the $\textit{whole}$ temperature range 140 MeV$<T<$162 MeV. 
 Note that for our two lightest quarks, the temperature range is 140 MeV$<T<$162 MeV.
 As seen from the left plot of Fig.~\ref{Fig:con}, the MEOS provides a good description of the subtracted chiral condensates in the whole temperature range for $m_{\pi}$ = 80, 90 MeV data sets.
 This good description suggests that the regular contribution to the chiral condensate is negligible for our two lightest quarks in the high temperature range.
We found that the $T_{c}$ obtained from a MEOS fit to $M$ in the whole temperature window 140 MeV$<T<$162 MeV is about 0.5 MeV larger than a MEOS fit in the temperature range 140 MeV$<T<$150 MeV.

It has been found in Ref.~\cite{Ding:2013lfa} that the scaling window in the case of HISQ action shrinks compared to the p4fat3 action on $N_{\tau}=$4 lattices~\cite{Ejiri:2009ac}.
To study the scaling behaviour in a pion mass window 80 MeV$<m_{\pi}<$160 MeV in a temperature range 140 MeV$<T<$150 MeV, it was sufficient to use a fit ansatz $f_{\mathrm{reg}}=(a_{0}+a_{1}\frac{T-T_{c}}{T_{c}})\frac{m_{l}}{m_{s}}$ to describe the chiral condensates via the O(2) MEOS fit to $M$~\cite{Ding:2015pmg}\cite{Ding:2013lfa}.
As can be seen from the middle plot of Fig.~\ref{Fig:con}, the deviation of the data points from the scaling curve in the high temperature region 150 MeV$<T<$170 MeV suggests that a higher order term in temperature is needed.
We thus perform a fit to the chiral condensate in our whole temperature range with a regular part $f_{\mathrm{reg}}$ including a higher order term in $T$,
\begin{small}
\begin{equation}
f_{\mathrm{reg}}=\left[ a_{0} + a_{1}\frac{T-T_{c}}{T_{c}} + a_{2}\left(\frac{T-T_{c}}{T_{c}}\right)^{2}\right]\frac{m_{{l}}}{m_{{s}}}.
\label{eq.fit}
\end{equation}
\end{small}
As seen from the right plot of Fig.~\ref{Fig:con}, the fit ansatz seems to be sufficient to describe the chiral condensates and this allows a description of the scaling violation at nonzero values of quark masses in the whole temperature range via the O(2) MEOS fit to the chiral order parameter $M$.
	
The critical temperature obtained from the MEOS fit with a regular term to $M$ in a temperature range 140 MeV$<T<$170 MeV is generally 1 MeV larger than the fit in a temperature range 140 MeV$<T<$150 MeV.
However, the obtained parameter $z_{0}$ agrees with each other within errors.
 
Although we have found that a good description of $M$ can be provided through O(2) scaling analyses,
we want to confirm that the physical point is above the tri-critical point from another perspective.
If $m_{s}^{\mathrm{phy}} < m_{s}^{\mathrm{tri}}$, it is expected that the chiral phase transition belongs to the Z(2) universality class. 
We then perform a fit to the chiral condensate at our two lowest quark masses using Eq.~(\ref{eq.M}) with the Z(2) scaling function and critical exponents.
Note that the breaking field now is $H=\frac{m_{l}}{m_{s}}-r_{c}$.
We also perform a Z(2) MEOS fit with a regular term as parameterized in Eq.~(\ref{eq.fit}) to the chiral condensate using all the data points.
The most important non-universal parameter $r_{c}$ determined from these scaling analyses turns out to be consistent with zero within errors, $r_c\lesssim 0.0001$, which corresponds to a bound on a possible critical
value for the Goldstone pion mass, $m_\pi^{crit} \lesssim 10$~MeV,
this suggests that $m_{s}^{\mathrm{tri}}<m_{s}^{\mathrm{phy}}$. 
\begin{figure}[ht]
	\begin{center}
		\includegraphics[width=0.43\linewidth]{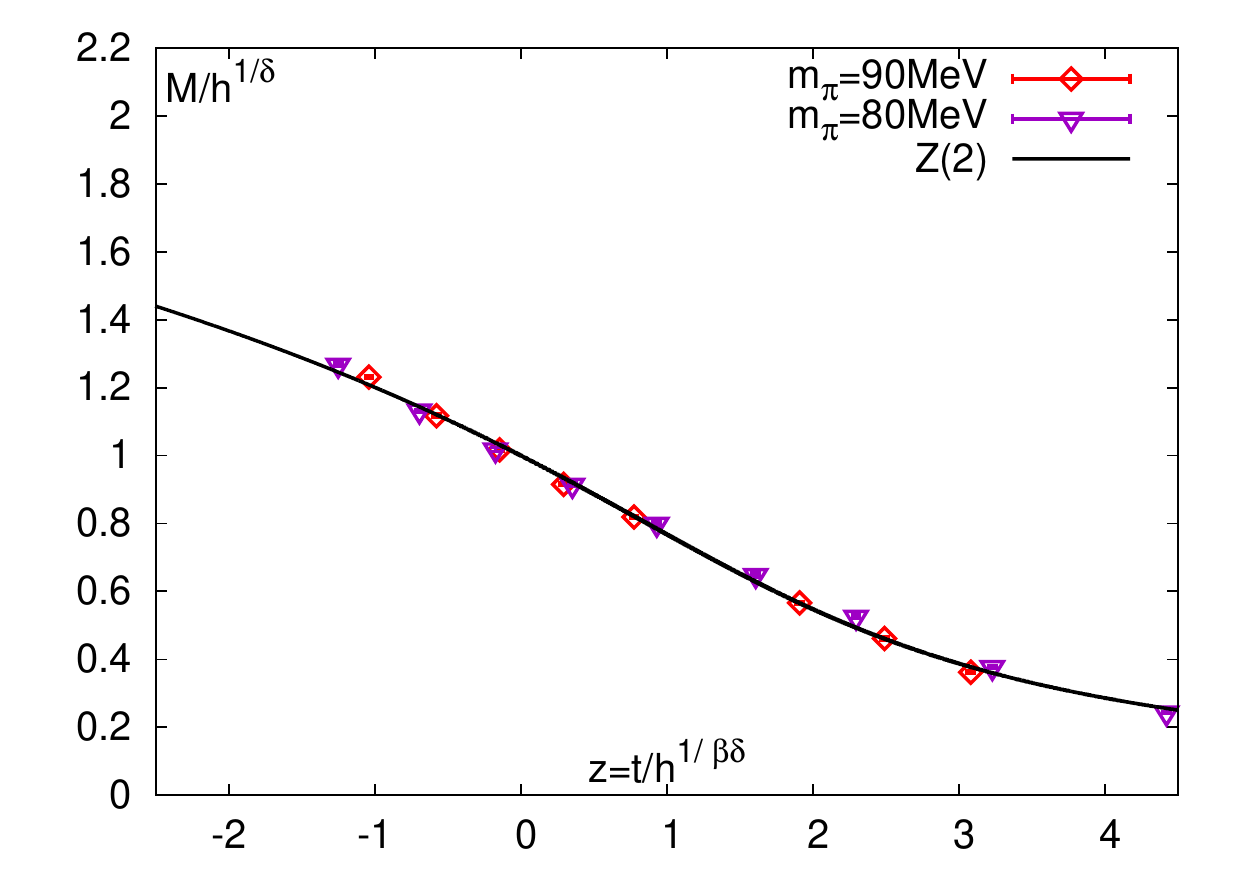}
		\includegraphics[width=0.43\linewidth]{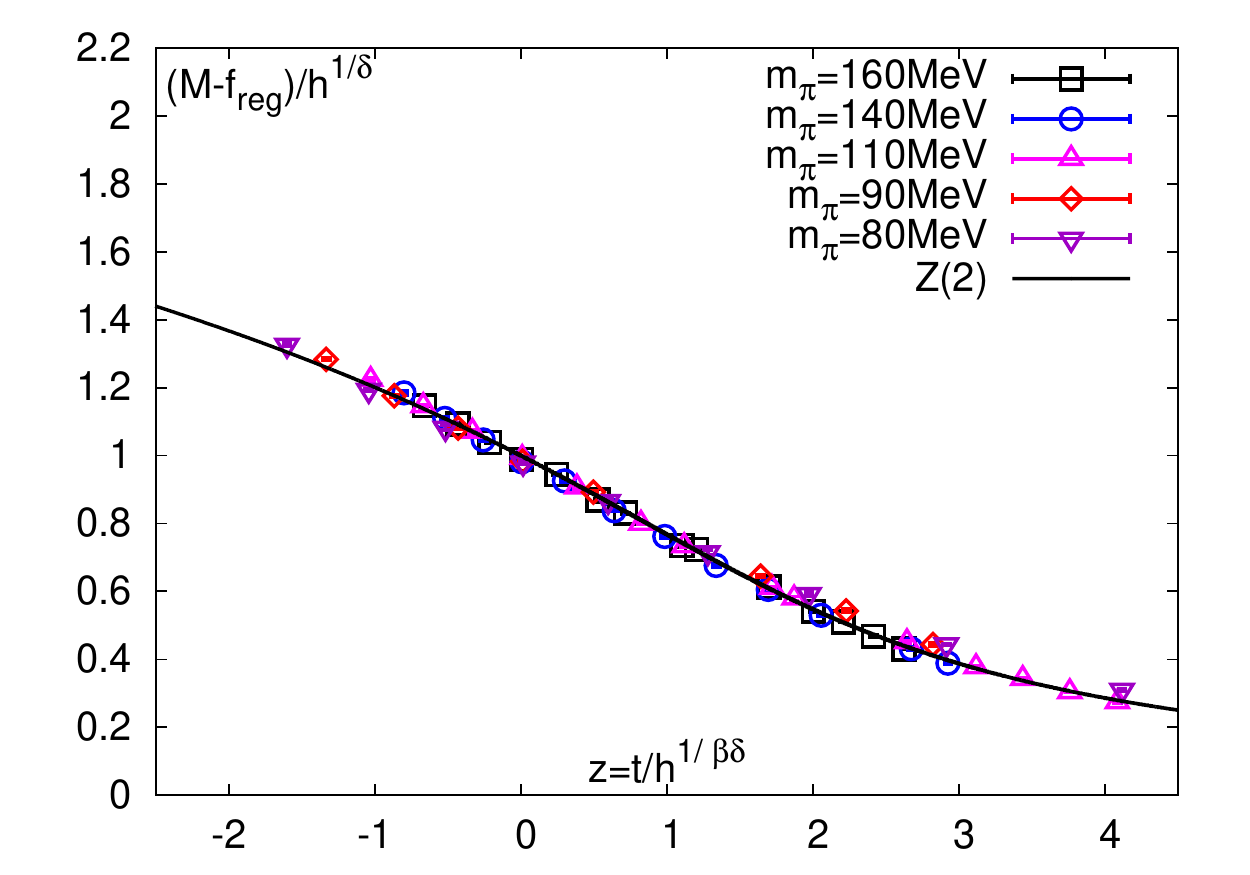}
	\end{center}
		\caption{Left: MEOS fit to $M$ using $m_{l}=m_{s}/80$ and
        $m_{l}=m_{s}/60$ data sets without a regular term in the
        temperature range 140 MeV$<T<$162 MeV. Right: MEOS fit to $M$
        using $\textit{all}$ the data sets with a regular term as
        expressed in Eq.~\protect\eqref{eq.fit} in the temperature range 140
        MeV$<T<$170 MeV.
    }
		\label{Fig:zcon}
\end{figure}

\section{Summary}
We studied the chiral phase structure in $N_{f}=$ 2+1 QCD in a larger temperature window 140 MeV$<T<$170 MeV compared to our previous study~\cite{Ding:2015pmg}.
The fit results suggested that the range of scaling variable z is linearly increasing with respect to the temperature range.
We found that the value of $T_{c}$ is not sensitive to the temperature region used in the MEOS analyses. 
Based on the O(2) and Z(2) universal scaling analyses on the subtracted chiral condensate, it suggests that $m_{s}^{\mathrm{tri}}<m_{s}^{\mathrm{phy}}$ and the system has a second order phase transition which, at non-zero values of the lattice spacing, belongs to the O(2) universality class in the chiral limit of light quarks in (2 + 1)-flavor QCD.

\end{document}